\documentclass[useAMS,usenatbib,usegraphicx]{mn2e}
\usepackage{lscape}
\newcommand\Mstar{M_\star}

\newcommand\zb{z_{\rm B}}

\newcommand\Mmin{M_{{\rm min}}}

\newcommand\Mh{\langle M_{h} \rangle}
\newcommand\Ng{\langle N_{g} \rangle}
\newcommand\Msun{{\rm M_{\odot}}}

\newcommand\sigmaM{\sigma_{\log M}}

\title[The very wide-field $gzK$ galaxy survey -- II]{The very wide-field $gzK$ galaxy survey -- II. The relationship between star-forming galaxies at $z \sim 2$ and their host haloes based upon HOD modelling}
\author[Ishikawa et al.]{Shogo~Ishikawa$^{1,2}$\thanks{E-mail: shogo.ishikawa@nao.ac.jp}, Nobunari~Kashikawa$^{1,2}$,  Takashi~Hamana$^{2}$, Jun~Toshikawa$^{2}$, 
\newauthor and Masafusa~Onoue$^{1,2}$\\
$^{1}$Depertment of Astronomy, School of Science, SOKENDAI (The Graduate University for Advanced Studies), Mitaka, Tokyo 181-8588, Japan\\
$^{2}$Optical and Infrared Astronomy Division, National Astronomical Observatory of Japan, Mitaka, Tokyo 181-8588, Japan}
\begin{document}

\pagerange{\pageref{firstpage}--\pageref{lastpage}} \pubyear{2015}

\maketitle

\label{firstpage}

\begin{abstract}
We present the results of an halo occupation distribution (HOD) analysis of star-forming galaxies at $z \sim 2$. 
We obtained high-quality angular correlation functions based on a large sgzK sample, which enabled us to carry out the HOD analysis. 
The mean halo mass and the HOD mass parameters are found to increase monotonically with increasing $K$-band magnitude, suggesting that more luminous galaxies reside in more massive dark haloes. 
The luminosity dependence of the HOD mass parameters was found to be the same as in the local Universe; however, the masses were larger than in the local Universe over all ranges of magnitude. 
This implies that galaxies at $z \sim 2$ tend to form in more massive dark haloes than in the local Universe, a process known as downsizing. 
By analysing the dark halo mass evolution using the extended Press--Schechter formalism and the number evolution of satellite galaxies in a dark halo, we find that faint Lyman break galaxies at $z \sim 4$ could evolve into the faintest sgzKs $(22.0 < K \leq 23.0)$ at $z \sim 2$ and into the Milky-Way-like galaxies or elliptical galaxies in the local Universe, whereas the most luminous sgzKs $(18.0 \leq K \leq 21.0)$ could evolve into the most massive systems in the local Universe. 
The stellar-to-halo mass ratio (SHMR) of the sgzKs was found to be consistent with the prediction of the model, except that the SHMR of the faintest sgzKs was smaller than the prediction at $z \sim 2$. 
This discrepancy may be attributed that our samples are confined to star-forming galaxies. 
\end{abstract}


\section{INTRODUCTION}
It is well known that galaxies are formed in highly dense regions of dark matter, termed dark haloes. 
According to the $\Lambda$CDM model, the large-scale structure of the Universe seen today originated from small density fluctuations of the dark matter in the early Universe. 
Galaxies are thought to have formed from cool gases trapped in the gravitational wells of the dark matter. 

Galaxy clustering is one of the most important keys to unveiling the galaxy evolution through cosmic time. 
The clustering amplitude of galaxies increases monotonically with the halo mass; i.e., more massive dark haloes are more strongly clustered. 
 
Many previous studies have revealed that galaxy clustering, which provides a measure of the dark halo mass, depends on the physical properties of galaxies; i.e., the brightness \citep{zehavi05}, the star-formation rate \citep[SFR;][]{bethermin14}, morphology \citep{zehavi02}, and colour \citep{zehavi02,zehavi05,zehavi11,coil08}. 
Because of recent galaxy selection methods that allow us to obtain large numbers of distant galaxies effectively, such as the dropout method \citep[e.g.,][]{madau96,steidel96} and the multi-colour selection method \citep[e.g.,][]{franx03,daddi04}, the clustering properties of the distant Universe have been described \citep[e.g,][]{kashik06,hilde09,jose13}, as well as those of the local Universe \citep[e.g.,][]{zehavi05,zheng07}. 
It is important to investigate the clustering evolution of galaxies with different physical properties to gain insight into galaxy evolution. 
Although galaxy formation reflects the distribution of dark matter, the complex baryonic processes that occur within dark haloes mean that galaxy evolution is not straightforward. 

The ``halo occupation distribution'' \citep[HOD, e.g.,][]{ma00,peacock00,seljak00,berlind02,berlind03} is a powerful theoretical approach that relates the galaxy distribution to the dark matter distribution. 
The HOD formalism describes the galaxy distribution within the host dark halo by characterizing them from the perspective of the probability distribution $P(N|M)$ that a halo of virial mass $M$ contains $N$ galaxies with specific physical properties, such as colour or galaxy type. 
One of the advantages of the HOD formalism is that the HOD parameters have explicit physical meanings; this enables us to easily interpret the relationship between the galaxies and the host haloes. 

The largest difference between the HOD framework and other galaxy--halo correspondence methods, such as the abundance matching method \citep[e.g.,][]{seljak00,cooray02}, is that the HOD model is based on a more realistic halo model and does not assume a one-to-one correspondence between galaxies and haloes. 
It is known that different types of galaxy pairs contribute to the two-point angular correlation function (ACF) between large-angular scales and small-angular scales. 
At the small-angular scale, galaxy clustering is contributed to mainly by galaxy pairs that are located in the same dark halo, referred to as the ``1-halo term'', whereas galaxy clustering at the large-angular scale is attributed to galaxy pairs that reside in different dark haloes, referred to as the ``2-halo term''. 
These two components result in different power-law slopes of the ACF at the two angular scales; these characteristics are well described using the HOD formalism. 

HOD analyses have been successful in revealing the clustering properties of galaxies and the characteristics of the host haloes both in the local Universe \citep[e.g.,][]{zehavi05,zehavi11,zheng07,zheng09,matsuoka11,coupon12} and the distant Universe \citep[e.g.,][]{hamana04,hamana06,jose13,bethermin14,durkalec14,mccracken15,martinez15}. 
\citet{zehavi05,zehavi11} carried out HOD analyses of the numerous local SDSS galaxies $(\sim 700,000$ galaxies over 8,000 deg$^{2})$ to investigate the luminosity and colour dependence of the clustering of galaxies. 
\citet{zehavi11} found that the clustering strength of the galaxies increased slowly at a luminosity of $L < L_{\ast}$ and increased rapidly at $L > L_{\ast}$. 
Moreover, \citet{zehavi11} found that the ACF of red galaxies has a strong clustering strength and steeper slope than that of blue galaxies, and faint red galaxies exhibit remarkably strong clustering at the small scale, which is comparable to the most luminous red galaxies. 
In the distant Universe, \citet{hamana06} calculated the mean dark halo mass and the expected number of galaxies within a dark halo at $z \sim 4$, and discussed an evolutionary relation between the Lyman break galaxies (LBGs) at $z \sim 4$ and the old passively evolving galaxies (OPEGs) at $z \sim 1$. 
Recently, \citet{durkalec14} implemented an HOD analysis of spectroscopically confirmed galaxies at $2.0 < z < 5.0$, and discussed galaxy evolution from high-$z$ to the local Universe. 
However, there have been few HOD analyses of $z \sim 2$ galaxies, due to difficulties in collecting a large sample of galaxies. 
Wide-field near infra-red (NIR) observations are required for a large sample of $z \sim 2$ galaxies; however, this is difficult because of the poor sky transparency and the limitations of the physical size of NIR detectors. 
\citet{bethermin14} carried out an HOD analysis on sBzK galaxies in the COSMOS field; however, the calculations of the dark halo mass were not sufficiently accurate due to the large errors of ACFs at large angular scales. 

The relationship between the spatial distribution of galaxies and the underlying invisible dark matter density field is known as galaxy bias, $b_{g}$. 
The HOD formalism can relate the dark halo and galaxy bias by comparing the ACFs determined from the assumed halo model and observation. 

Another method that can be used to relate the dark halo to the distribution of baryons in the dark halo is termed the ``stellar-to-halo mass ratio'' (SHMR). 
The SHMR has been the subject of many studies aiming to reveal or constrain the properties of galaxies and the conversion efficiency from baryons to galaxies in dark haloes \citep[e.g.,][]{behroozi10,behroozi13,moster10,yang12,leau12}. 
On the $M_{{\rm h}}$ versus $\Mstar/M_{{\rm h}}$ diagram, the SHMR has a peak at $M_{{\rm h}} \sim 10^{12}$ $\Msun$, which corresponds to the most efficient mass (pivot halo mass) to convert from baryonic matter to stars, which is consistent with theoretical prediction \citep[e.g.,][]{blu84}; however, SHMR at the peak halo mass is much smaller than the universal baryon fraction $(\Omega_{{\rm b}} / \Omega_{{\rm m}} \sim 0.167)$ \citep[e.g.,][]{behroozi10,leau12,yang12}. 
This characteristic dark halo mass is determined by the equilibrium of the suppression mechanism that star-formation becomes equal between the shallow gravitational potential wells for less-massive dark haloes and AGN feedback, and high virial temperature of massive dark haloes. 
\citet{leau12} showed that this characteristic dark halo mass varies with the redshift, which is evidence of mass downsizing at $0 < z < 1$. 
\citet{behroozi13} predicted the redshift evolution of the SHMR from $z = 8$ to $z = 0$ using numerical simulations. 
They concluded that star formation was most efficient at $M_{{\rm h}} \sim 10^{12}$ $\Msun$ at $0 < z < 5$; however, there was a trend that the pivot halo mass became larger with increasing redshift from $z =0$ to $z = 3$, and the conversion efficiency for the massive dark halo was the highest when $ 2 < z < 3$. 
The SHMR has been calculated based on observations at low-$z$ and $z \sim 3$; however, there have been few investigations at $z \sim 2$ because of the lack of estimates of the dark halo mass. 

In this work, we describe an HOD analysis of a large number of sgzKs, exploiting high-quality ACFs to investigate the properties of dark haloes at $z \sim 2$, as well as the relationship between the sgzKs and other galaxy populations. 
The HOD formalism is based upon a halo model, and it enables us to calculate the dark halo mass precisely, as well as to determine the distribution of sgzKs in their host dark haloes; this enables us to connect sgzKs to their ancestors/descendants. 

The aim of this paper is to discuss galaxy evolution from high-$z$ to the local Universe from the point of view of the evolution of the dark halo mass and the number of satellite galaxies. 
The framework of this paper is organized as follows. 
In Section~2, we summarize previous research, and discuss the properties of the sgzK galaxy sample and its clustering properties. 
The details of our HOD model and the HOD analysis are described in Section~3, and the results of HOD analysis are given in Section~4. 
In Section~5, we discuss the evolution of dark halo mass, the number of satellite galaxies, and the relationship between the dark halo mass and the stellar mass of sgzKs using the results of HOD analysis. 
Conclusions are drawn in Section~6. 
All magnitudes and colours are in the AB system. 
Throughout this paper, we assume the flat lambda cosmology $(\Omega_{\rm m} = 0.3$, $\Omega_{\Lambda} = 0.7)$, the Hubble constant as $h = {\rm H_{0}} / 100 \, {\rm km \, s^{-1} \, Mpc^{-1}} = 0.7$, and the matter fluctuation amplitude at $8 h^{-1}$Mpc as $\sigma_{8} = 0.8$. 
With these cosmological parameters, the age of the Universe at $z \sim 2$ is $\sim 3.22$ Gyr and 1 arcsec corresponds to $8.73h^{-1}$ kpc in the comoving scale. 

\section{THE sgzK GALAXY SAMPLE AND CLUSTERING ANALYSIS}
In our previous study \citep[][hereafter referred to as paper~I]{ishikawa15}, we addressed difficulties in constructing a large sample of galaxies at $z \sim 2$ using both our data and publicly available data, and constructed the largest star-forming galaxy sample yet at $z \sim 2$. 
In this section, we briefly review our data and the clustering analysis described in paper~I. 

We retrieved the $K$-band image public archive data on the VIMOS4 region from the United Kingdom Infra-Red Telescope (UKIRT) Deep Sky Survey \citep[UKIDSS;][]{law07} in the Deep Extragalactic Survey (DXS). 
We obtained $\zb$-band photometric data for $\sim$ 3 deg$^{2}$ \citep{kashik15} for the region, and retrieved Suprime-Cam $z$-band image data from the SMOKA data archive server \citep{baba02} to extend the survey field to $5.2$ deg$^{2}$. 
The $\zb$-band ($\lambda_{{\rm c}} = 8,842$${\rm \AA}$, ${\rm FWHM} = 689$${\rm \AA}$) is a custom-made filter~\citep{shima05}. 
We retrieved $g$-band images, which cover the entire field where $K$- and $\zb/z$-band images were available, using the Canada--France--Hawaii Telescope Legacy Survey \citep[CFHTLS;][]{gwyn11} archive data, rather than using $B$-band data. 
The limiting magnitudes of these $K$-, $\zb /z$-, and $g$-band data were $K=23.31$ ($3\sigma$, $2''$ in AB magnitude), $\zb = 25.55$, $z = 25.36$, and $g = 26.08$, respectively. 
The limiting magnitude of our $K$-selected catalogue was $K = 23.0$, which corresponds to a 70$\%$ completeness cut. 

Differences in the filter set used for our $g\zb/zK$ and the original VLT-$BzK$ filters should be considered when selecting galaxies, to ensure the same properties as in previous BzK studies. 
We carefully implemented band corrections for our $(g - \zb)$, $(g -z)$, $(\zb -K)$, and $(z - K)$ colours to convert them into VLT-$(B -z)$ and -$(z -K)$ colours (see Section~2.3 of paper~I for further details). 
Following these corrections, we applied the BzK selection method (termed here the ``gzK selection'' method) to obtain sgzKs. 
The original BzK selection criterion proposed by \citet{daddi04} was applied, as we may consider that the colours of the objects listed in our $K$-selected catalogue have the same values as those colours observed using the VLT. 
The criterion used to select sgzKs is given by 
\begin{equation}
(z -K) \geq (B-z) -0.2. 
\label{eq:daddi_sgzK}
\end{equation}
The number of selected sgzKs was $41,112$, which represents the largest sgzK/sBzK galaxy sample yet obtained. 
The number counts of all $K$-selected galaxies and sgzKs were in good agreement with the results of previous studies. 

The derived ACF amplitude exhibited a clear $K$-band luminosity dependence, showing that more luminous sgzKs were more strongly clustered than fainter sgzKs; this trend had been inferred by previous studies \citep[e.g.,][]{hayashi07}. 
Moreover, with our ACFs, an excess from a power law at the small angular scale was clearly seen because of the high signal-to-noise (S/N) ratio, due to the large number of samples. 

Our correlation lengths showed excellent agreement with the results of almost all previous studies over all magnitude ranges. 
There was a tendency for bright sgzKs to exhibit stronger clustering than faint sgzKs, as was the case with the amplitude of the ACFs. 
We also investigated the relationship between clustering strength and the stellar mass of the sgzKs, which was consistent with previous studies \citep{bethermin14,bielby14}. 

The dark halo masses were calculated by comparing the effective bias derived from the correlation length with the halo model based on \citet{sheth01b} and \citet{mowhite02}. 
The mean dark halo masses of sgzKs were found to be $(1.32^{+0.09}_{-0.12}) \times 10^{12}$ $h^{-1} \Msun$, $(2.55^{+0.34}_{-0.32}) \times 10^{12}$ $h^{-1} \Msun$, and $(3.26^{+1.23}_{-1.02}) \times 10^{13}$ $h^{-1} \Msun$ for $22.0 < K \leq 23.0$, $21.0 < K \leq 22.0$ and $18.0 \leq K \leq 21.0$, respectively. 
This suggests that more luminous sgzKs in the $K$-band reside in more massive dark haloes. 
We also found that the relationship between the minimum mass $\Mmin$ and the mean mass $\Mh$ of the dark halo satisfied $\Mh \approx 3\Mmin$, which is consistent with the data reported by \citet{hayashi07}. 

\section{HOD ANALYSIS}
\subsection{The HOD Model}
Some methods used to calculate the dark halo mass, such as the ACF and abundance matching, assume that one dark halo must possess only one galaxy; however, this is not always the case, especially for more massive dark haloes, which often contain satellite galaxies. 
Galaxy clusters, which are embedded in large massive dark haloes, have numerous member galaxies, whereas less massive dark haloes may contain no galaxy at all. 
Thus, a more precise halo model is required to describe the distribution of galaxies within dark haloes. 

The HOD model was proposed to implement more detailed analysis by linking virialized dark haloes and the distribution of galaxies within them. 
The HOD framework provides a probability distribution for the number of galaxies in the dark halo as a function of the dark halo mass. 
In this paper, we used a standard HOD model proposed by \citet{zheng05}, who confirmed  the validity of their model by comparing observational results with a smoothed particle hydrodynamics (SPH) simulation and a semianalytic galaxy formation model. 
In this HOD model, galaxy occupation is clearly separated by the contribution of central galaxy, $N_{c}(M)$, and satellite galaxies, $N_{s}(M)$, and thus total galaxy occupation within dark halo, $N(M)$, is described as
\begin{equation}
N(M) = N_{c}(M) \times \bigl[1 + N_{s}(M) \bigr]. 
\label{eq:N_gal}
\end{equation}
The central galaxy occupation that satisfies the magnitude threshold is characterised by a step-like function with smoothed-cut-off as
\begin{equation}
N_{c}(M) = \frac{1}{2} \Bigl[1 + {\rm erf} \Bigl(\frac{\log M - \log {M_{{\rm min}}}}{\sigma_{\log M}} \Bigr) \Bigr], 
\label{eq:N_cent}
\end{equation}
whereas the satellite galaxy occupation that satisfies the magnitude threshold is described by a Poisson distribution; i.e., 
\begin{equation}
N_{s}(M) = \Bigl(\frac{M - M_{0}}{M_{1}} \Bigr)^{\alpha}. 
\label{eq:N_sate}
\end{equation}
We note that we set $N_{s}(M) = 0$ if the dark halo mass $M$ satisfies $M < M_{0}$. 

We have five free HOD parameters $\Mmin$, $M_{1}$, $M_{0}$, $\alpha$, and $\sigmaM$. 
One of the advantages of the HOD model is that these HOD parameters have explicit physical meanings. 
$\Mmin$ is an approximation of the minimum dark halo mass required to possess a central galaxy within a dark halo and $\sigmaM$ is the cut-off parameter of central galaxy occupation. 
For $N_{s}(M)$, $M_{0}$ is the minimum dark halo mass to form a satellite galaxy within a dark halo, $M_{1}$ is the typical dark halo mass that possesses one satellite galaxy, and $\alpha$ is the power-law slope of the relationship between dark halo mass and the number of satellite galaxies. 

The HOD framework allows for the formation of multiple galaxies within a single dark halo, which can explain the excess from a power law in the observed ACFs at the small angular scale (1-halo term), corresponding to less than the virial radius. 
At the large angular scale, especially at scales larger than the virial radius (2-halo term), the ACF is contributed to by galaxy pairs residing in different dark haloes; i.e., the ACF becomes the simple power law. 

For the computation of ACFs in the HOD framework, we assumed the halo mass function of \citet{sheth99}, halo bias of \citet{tinker05}, and dark halo profile of an NFW profile \citep{nfw}. It is noted that the HOD analysis also requires the redshift distributions of each subsample. 
We derived the redshift distributions, which are also required in the HOD analysis of each sgzK subsample using the COSMOS $K$-selected catalogue \citep{muzzin13}. 
Further details of these processes and the derived redshift distribution of each subsample are found in paper~I. 

\subsection{HOD Analysis on the sgzKs}
The HOD parameters can be determined by fitting the HOD 1-halo/2-halo ACFs to the observed ACFs using a least $\chi^{2}$ method; i.e., 
\begin{eqnarray}
\chi^{2} & = & \sum_{i, j} \Bigl[\omega^{\rm obs}(\theta_{i}) - \omega^{\rm HOD}(\theta_{i}) \Bigr](C^{-1})_{ij}  \Bigl[\omega^{\rm obs}(\theta_{j}) - \omega^{\rm HOD}(\theta_{j}) \Bigr] \nonumber \\
& + & \frac{[n_{g}^{{\rm obs}} - n_{g}^{{\rm HOD}}]^{2}}{\sigma^{2}_{n_{g}}}, 
\label{eq:sqchi_HOD}
\end{eqnarray}
where $\omega^{\rm HOD}(\theta_{i})$ and $\omega^{\rm obs}(\theta_{i})$ are the model-predicted and the observed ACFs at a given angular scale, $n_{g}$ is the galaxy number density, and $\sigma^{2}_{n_{g}}$ is the statistical 1$\sigma$ errors of the observed number densities of sgzK galaxies, respectively. 
We note that the error of the galaxy number density, $\sigma_{n_{g}}$, is considered as both a Poisson error and cosmic variance \citep{torenti08}. 
$(C^{-1})_{ij}$ is the $(i, j)$ element of the inverse covariance matrix calculated by the Jackknife method (see below for more details). 
Galaxy number density from the HOD model is calculated by integrating the halo mass function, by weighting the number of galaxies within dark haloes (equation \ref{eq:N_gal}) as
\begin{equation}
n_{g}^{{\rm HOD}} = \int dM n(M) N(M), 
\label{eq:galdens}
\end{equation}
where $n(M)$ is the halo mass function proposed by \citet{sheth99}. 

\begin{figure*}
\centering
\includegraphics[width=168mm]{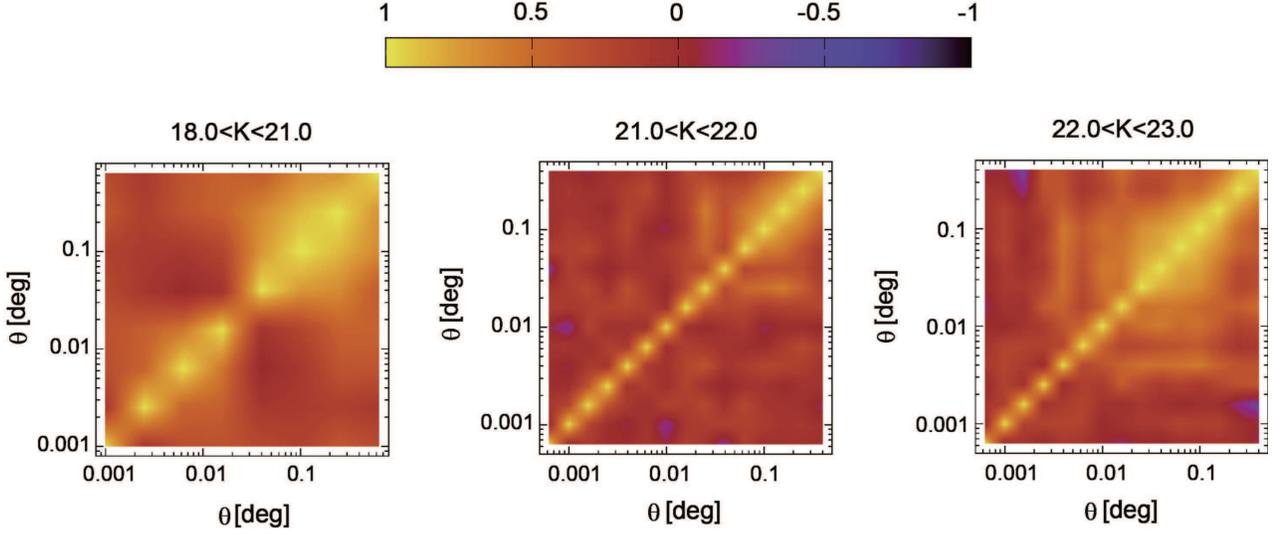}
\caption{Correlation coefficient of the sgzK galaxy sample. Left, middle, and right panels show the correlation coefficient results of $18.0 \leq K \leq 21.0$, $21.0 < K \leq 22.0$, and $22.0 < K \leq 23.0$ subsamples, respectively. These results were derived using the Jackknife resampling method. }
\label{fig:corr_coeff}
\end{figure*}

It is useful to calculate the covariance matrix to estimate the correlation between other angular bins. 
We calculated the covariance matrix, $C_{ij}$, of our sgzK galaxy sample using the Jackknife resampling method. 
We divided our survey region into $N=64$ and calculated the ACF by removing one sub-region, repeating this process 64 times. 
The covariance matrix is calculated by 
\begin{equation}
C_{ij} = \frac{N-1}{N} \sum_{k=1}^{N} \bigl(\omega_{k}(\theta_{i}) - \overline{\omega}(\theta_{i}) \bigr) \times \bigl(\omega_{k}(\theta_{j}) - \overline{\omega}(\theta_{j}) \bigr), 
\label{eq:cm}
\end{equation}
where $\overline{\omega}(\theta_{i})$ is the average ACF of $i-$th angular bin. 
The correlation factor of \citet{hartlap07} is applied in the case that we convert our covariance matrix into an inverse matrix. 
As \citet{coupon12}, we show the correlation coefficient of our sgzK galaxy sample, $r_{ij}~=~C_{ij} / \sqrt{C_{ii}C_{jj}}$, for each luminosity subsample (Figure~\ref{fig:corr_coeff}). 

$\omega^{\rm HOD}$ is composed of two components, which are the HOD derived ACFs of 1-halo/2-halo terms; i.e., 
\begin{equation}
\omega^{\rm HOD}(\theta_{i}) = \omega_{\rm 1h}(\theta_{i}) + \omega_{\rm 2h}(\theta_{i}), 
\end{equation}
where $\omega_{\rm 1h}(\theta)$ and $\omega_{\rm 2h}(\theta)$ are the ACFs of the 1-halo term and 2-halo term, respectively. 
We fitted $\omega^{\rm HOD}$ by varying the five free HOD parameters $(\Mmin$, $M_{1}$, $M_{0}$, $\alpha$, and $\sigmaM$), using the ``Population Monte Carlo (PMC)'' technique \citep{cappe07}. 
The publicly available code, ``CosmoPMC'', was adopted to investigate the best-fit HOD parameters in parameter space \citep{wraith09,kilbinger10}. 
First, we estimated best-fit HOD parameters of the overall sgzKs $(18.0 \leq K \leq 23.0)$ by varying the five parameters over a wide range. 
Then, differentially resampled ACFs were used to investigate the dependence of the HOD parameters on $K$-band luminosity. 
We carried out HOD analysis in two ways: (1) by varying the five HOD parameters and (2) by only varying $\Mmin$, $M_{1}$, and $M_{0}$ with fixed $\alpha=1.19$ and $\sigmaM=0.254$ corresponding to the best-fit values for the total sample. 

As in \citet{hamana06} and \citet{bethermin14}, we define the mean halo mass, $\Mh$, and the expectation number of galaxies, $\Ng$, weighted by the number of galaxies within the dark halo as follows: 
\begin{equation}
\Mh  =  \frac{\int_{\Mmin}^{\infty} dM \, M n(M) N(M)}{\int_{\Mmin}^{\infty} dM \, n(M) N(M)} 
\label{eq:Mhalo}
\end{equation}
and
\begin{equation}
\Ng  =  \frac{\int_{\Mmin}^{\infty} dM \, n(M) N(M)}{\int_{\Mmin}^{\infty} dM \, n(M)}, 
\label{eq:Ng}
\end{equation}

where the low-mass cut-off of the integration was set to $\Mmin$ to exclude the contribution of low-mass dark haloes with no galaxies. 

\section{RESULTS OF THE HOD ANALYSIS}
\subsection{Fitting Parameters}
\begin{figure*}
\centering
\includegraphics[width=168mm]{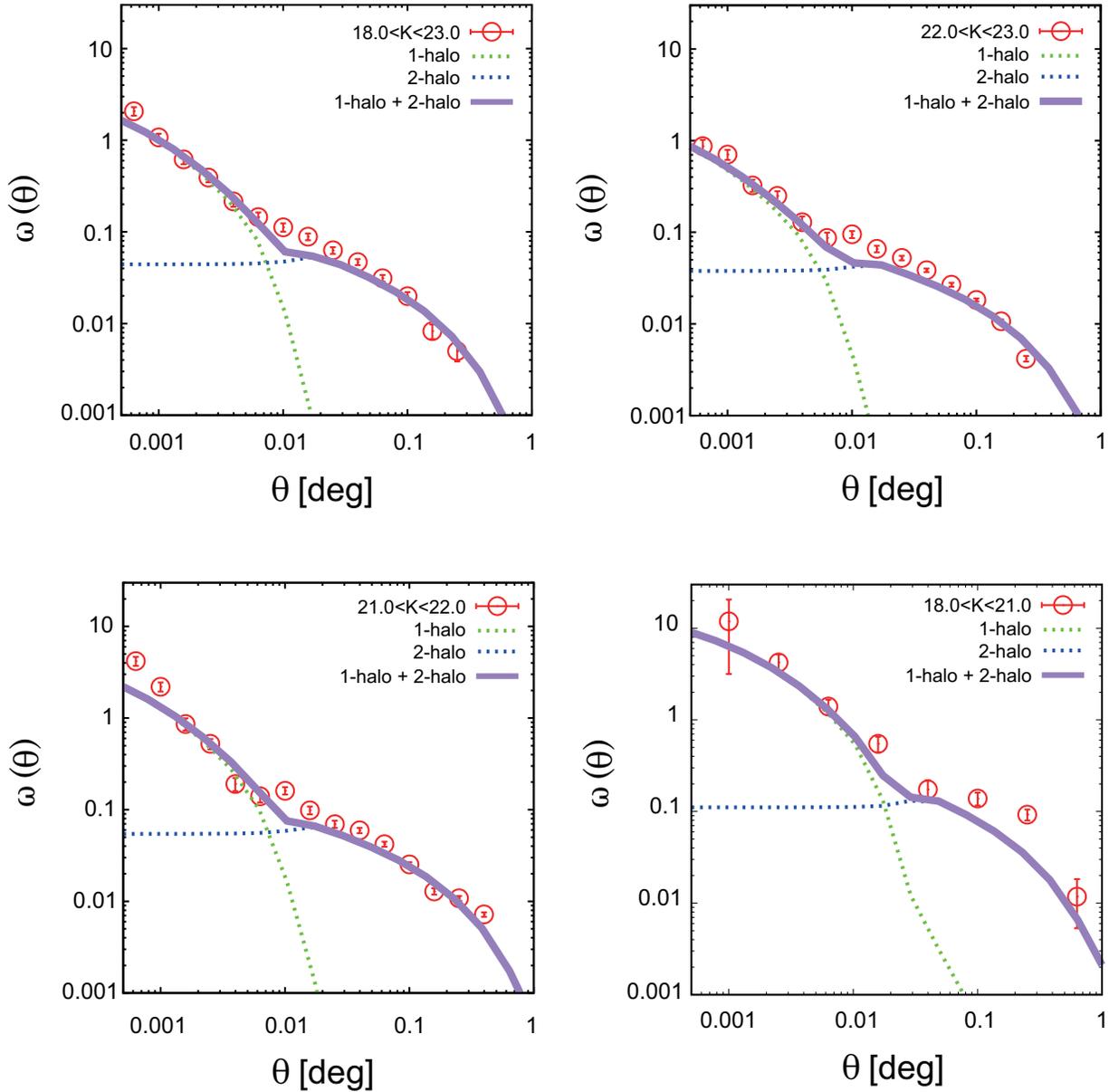}
\caption{A comparison between the observed ACF (red circles) and the best-fit HOD ACF. The dashed green (blue) curve represents the ACF of 1-halo (2-halo) term, and the solid purple curve shows a best-fit HOD ACF composed of the sum of the 1-halo term and the 2-halo term. We note that the full sample result (top-left) was derived by varying the five HOD parameters (see text for more details); the and results for the differential magnitude-cut subsamples shown in the other panels were derived by fixing $\alpha$ and $\sigmaM$. }
\label{fig:HOD_result}
\end{figure*}

Figure \ref{fig:HOD_result} shows the best-fit HOD models to the observed ACFs of the sgzK. 
It is clear that our HOD model can describe the observed ACFs of sgzKs. 
The best-fit HOD parameters with $1\sigma$ error are listed in Table \ref{tab:HOD_param}. 
We show the results of the best-fit HOD parameters for the case varying all HOD parameters, as well as the case with fixed $\alpha$ and $\sigmaM$. 
Fixed $\alpha$ and $\sigmaM$ are derived from the HOD analysis of the total sgzK (satisfying $18.0 \leq K \leq 23.0$) samples. 
Some studies report the best-fit values of $\alpha$ and $\sigmaM$ of various galaxy populations, although most of the HOD analysis studies used a fixed $\alpha$ and/or $\sigmaM$. 

Most studies support our fixed $\alpha$ and $\sigmaM$ \citep[e.g.,][]{zehavi05,coupon12}. 
\citet{martinez15} carried out HOD analysis on $Spitzer$-selected galaxies at $z = 1.5$ by fixing $\alpha = 1.0$ and $\sigmaM = 0.2$, which are consistent with our fixed $\alpha$ and $\sigmaM$. 
In the local Universe, \citet{zehavi11} also showed $\alpha \sim 1$ and $\sigmaM \sim 0.2$ for their faint and intermediate galaxy sample $(M_{r}^{max} < -20.5)$, although the result of their full sample, including the brightest galaxies $(M_{r}^{max} < -22.0)$, was $\sigmaM \sim 0.7$. 
From here, we adopt the results of the HOD analysis in which $\alpha$ and $\sigmaM$ are fixed. 

At $\sim 0.01$ deg, there are systematic spikes in our ACFs, whose characteristics cannot be explained by the HOD model. 
The origin of this discrepancy is unclear. 
One possible explanation is that the halo model cannot describe the observed ACF at the scale of the virial radius. 
Several previous studies, especially at $z \sim 2$, also showed the small excess of the observed ACF from the best-fit HOD at this scale \citep[e.g.,][]{wake11,martinez15,mccracken15}; however, further discussion is beyond the scope of this study. 

\begin{figure*}
\centering
\includegraphics[width=168mm]{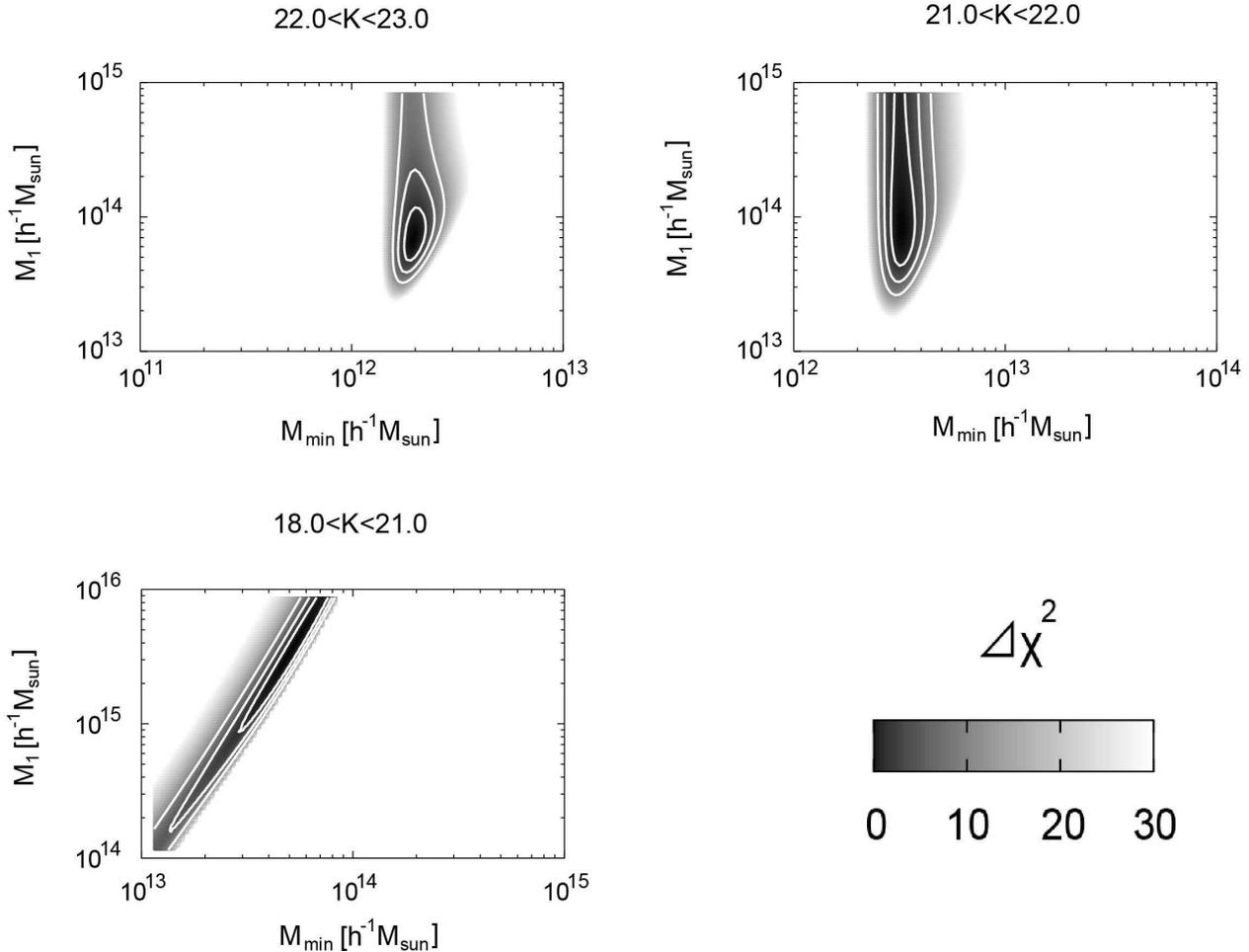}
\caption{Confidence contour maps of the HOD parameters derived from the least $\chi^{2}$ method on the $M_{1}$ -- $\Mmin$ parameter planes, fixing $M_{0}$, $\sigmaM$, and $\alpha$ of the best-fit values. 
The grey-scale indicates the difference in $\chi^{2}$ from the value of the best-fit parameters. 
The contour lines represent 1$\sigma$, 2$\sigma$, and 3$\sigma$ confidence levels. }
\label{fig:contour}
\end{figure*}

\begin{figure}
\centering
\includegraphics[width=84mm]{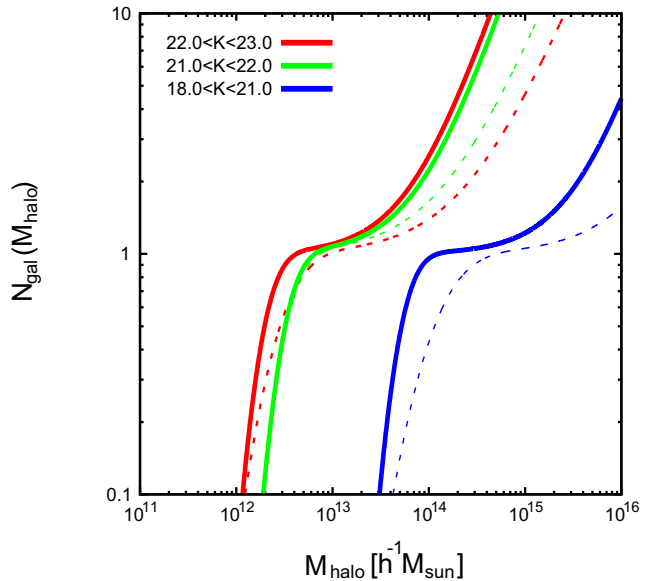}
\caption{The halo occupation functions of the best-fit HOD models of each sgzK subsample. We show both results of varying all free parameters (dashed lines) and fixing $\alpha$ and $\sigmaM$ (solid lines). }
\label{fig:hof}
\end{figure}

The errors of the mean halo mass, $\Mh$, and the expectation number of galaxies, $\Ng$, were estimated from the minimum/maximum values of $\Mh$ and $\Ng$ by varying the free parameter set. 
Figure~\ref{fig:contour} shows $\chi^{2}$ confidence contour maps of the HOD parameters derived from the least $\chi^{2}$ method on the $M_{1}$--$\Mmin$ parameter planes fixing $M_{0}$, $\sigmaM$, and $\alpha$ of those best-fit values. 
As shown by the data listed in Table \ref{tab:HOD_param}, the mass quantities $M_{1}$, $\Mmin$, $M_{0}$, and $\Mh$ almost increased as the magnitude thresholds became brighter. 
This is evidence that brighter sgzKs reside in more massive dark haloes. 
Figure \ref{fig:hof} shows the halo occupation functions (equation~\ref{eq:N_gal}) of each luminosity subsample as a function of the dark halo mass. 
We show both halo occupation functions in the case of varying five HOD parameters (dashed lines) and fixing $\alpha$ and $\sigmaM$ (solid lines). 

The expectation number of galaxies within a dark halo, $\Ng$, was similar between each subsample. 
It follows that the ratio of the number of $K$-faint sgzKs in less-massive dark haloes to the number of $K$-bright sgzKs in massive dark haloes was close to unity. 
Nevertheless, we should keep in mind that $\Ng$ is strongly affected by $\alpha$, which represents the satellite galaxy formation efficiency in dark haloes, and which was fixed in this analysis. 
A possible dependence of the galaxy limiting magnitude on $\alpha$ should be investigated using a larger galaxy sample as part of a future extensive survey, such as the Hyper Suprime-Cam (HSC) survey. 

\begin{table*}
\centering
\begin{minipage}{180mm}
\caption{The best-fit HOD parameters. }
\scalebox{0.80} {
\begin{tabular}{lcccccccc}
\hline\hline
magnitude threshold & $M_{1}$ $[h^{-1} \Msun]$& $\Mmin$ $[h^{-1} \Msun]$& $M_{0}$ $[h^{-1} \Msun]$ \footnote[1]{The number of the brightest sgzK was too small to put constraints on the 1$\sigma$ limit for $M_{0}$. }&  $\sigmaM$ & $\alpha$ & $\Mh$ $[h^{-1} \Msun]$ &  $\Ng$ & $\chi^{2}/{\rm dof}$ \\ \hline
$22.0 < K \leq 23.0$ & $(2.62^{+0.35}_{-0.25}) \times 10^{14}$ & $(2.88^{+0.08}_{-0.08}) \times 10^{12}$ & $(4.10^{+11.8}_{-2.89}) \times 10^{10}$ & $0.404^{+0.095}_{-0.132}$ & $0.961^{+0.125}_{-0.075}$ & $(5.00^{+1.13}_{-0.85}) \times 10^{12}$ & $0.788^{+0.080}_{-0.051}$ & 1.57 \\
$21.0 < K \leq 22.0$ & $(1.53^{+0.97}_{-0.31}) \times 10^{14}$ & $(3.37^{+0.11}_{-0.10}) \times 10^{12}$ & $(1.80^{+6.27}_{-1.40}) \times 10^{11}$ & $0.292^{+0.243}_{-0.122}$ & $0.983^{+0.141}_{-0.150}$ & $(6.61^{+1.14}_{-2.33}) \times 10^{12}$ & $0.855^{+0.125}_{-0.127}$ & $2.71$ \\
$18.0 \leq K \leq 21.0$ & $(1.75^{+0.60}_{-1.34}) \times 10^{16}$ &  $(1.15^{+0.76}_{-0.43}) \times 10^{14}$ & $4.45 \times 10^{9}$ & $0.477^{+0.165}_{-0.531}$ & $0.993^{+0.262}_{-0.155}$ & $(5.89^{+8.26}_{-3.13}) \times 10^{13} $ & $0.638^{+0.084}_{-0.056}$ & $2.46$ \\ \hline
$22.0 < K \leq 23.0$ & $(6.92^{+0.81}_{-0.65}) \times 10^{13}$ & $(1.98^{+0.05}_{-0.06}) \times 10^{12}$ & $(3.92^{+13.5}_{-2.93}) \times 10^{10}$ & $0.254 \, {\rm (fixed)}$ &  $1.19 \, {\rm (fixed)}$  & $(4.55^{+0.42}_{-0.60}) \times 10^{12}$ & $0.877^{+0.071}_{-0.064}$ & 3.02 \\
$21.0 < K \leq 22.0$ & $(8.39^{+3.11}_{-1.76}) \times 10^{13}$ & $(3.25^{+0.11}_{-0.10}) \times 10^{12}$ & $(1.91^{+11.9}_{-1.65}) \times 10^{11}$ & $0.254 \, {\rm (fixed)}$ &  $1.19 \, {\rm (fixed)}$ & $(6.91^{+0.77}_{-1.05}) \times 10^{12}$ & $0.883^{+0.095}_{-0.084}$ & $2.64$ \\
$18.0 \leq K \leq 21.0$ & $(3.55^{+6.13}_{-2.23}) \times 10^{15}$ & $(5.22^{+2.24}_{-1.42}) \times 10^{13}$ & $4.03 \times 10^{9}$ & $0.254 \, {\rm (fixed)}$ & $1.19 \, {\rm (fixed)}$ & $(6.12^{+1.92}_{-1.37}) \times 10^{13} $ & $0.759^{+0.028}_{-0.022}$ & $2.52$ \\
\hline\hline
\label{tab:HOD_param}
\end{tabular}
}
\end{minipage}
\end{table*}

\subsection{Dark Halo Mass of sgzK Galaxies}
In paper~I, we determined the dark halo mass corresponding to sgzKs using large-scale galaxy clustering \citep{mo96}, which assumes a one-to-one correspondence. 
HOD analysis enables us to derive a more accurate dark halo mass based on a more realistic HOD halo model. 

\begin{figure}
\centering
\includegraphics[width=84mm]{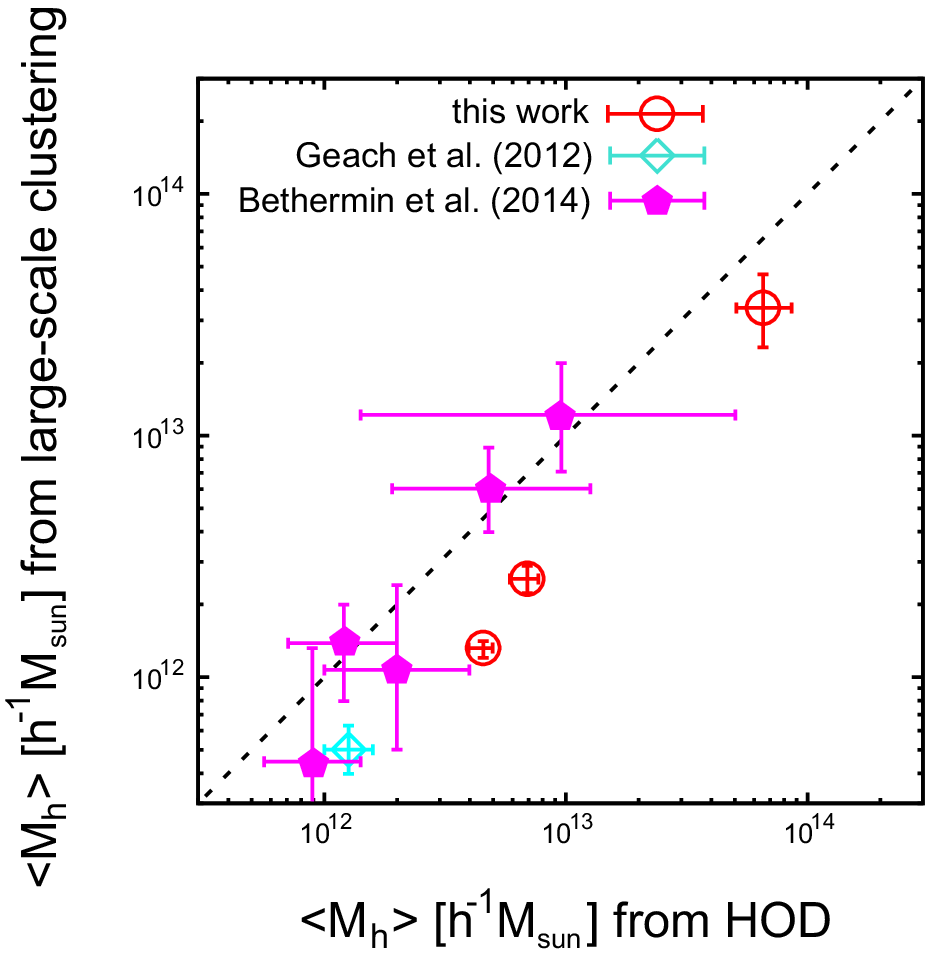}
\caption{A comparison of the dark halo masses calculated from the HOD model and large-scale galaxy clustering. The dotted-black line represents the one-to-one relationship between halo masses derived by both methods. Note that the results reported by \citet{bethermin14} were the dark halo masses of stellar-mass-limited subsamples. }
\label{fig:comp_acf-hod}
\end{figure}

Figure \ref{fig:comp_acf-hod} shows a comparison of the dark halo masses calculated using the HOD analysis and the ACFs. 
The dark halo masses calculated using the HOD analysis were several times larger than those calculated using the ACFs. 
We note that the definitions of the mean halo masses with these two methods differ slightly. 
The mean halo mass from the HOD analysis is weighted by the number of galaxies in a dark halo, which tends to make $\Mh$ larger than the estimate from the ACF; that is because HOD formalism does not assume a one-to-one correspondence, as also mentioned in \citet{geach12}. 

We also compared our sgzK dark halo masses with those of the previous studies, as shown in Figure \ref{fig:comp_acf-hod}. 
\citet{geach12} investigated the mass of H$\alpha$ emitters (HAEs) at $z = 2.23$, and \citet{bethermin14} estimated the mass of sBzKs at $z \sim 2$ in the COSMOS field. 
Our results are in good agreement with these data. 
\citet{bethermin14} reported little difference between data obtained using the two methods. 
However, this may be caused by large error bars in their HOD analysis. 

\section{DISCUSSION}
\subsection{The Luminosity Dependence of the HOD Parameters}
As we showed in Section~4.1, the HOD parameters exhibit a dependence on the $K$-band luminosity, which has been reported previously \citep[e.g.,][]{zehavi11,martinez15}. 
This is caused by the fact that more-luminous galaxies generally reside in more-massive dark haloes. 

\begin{figure}
\centering
\includegraphics[width=84mm]{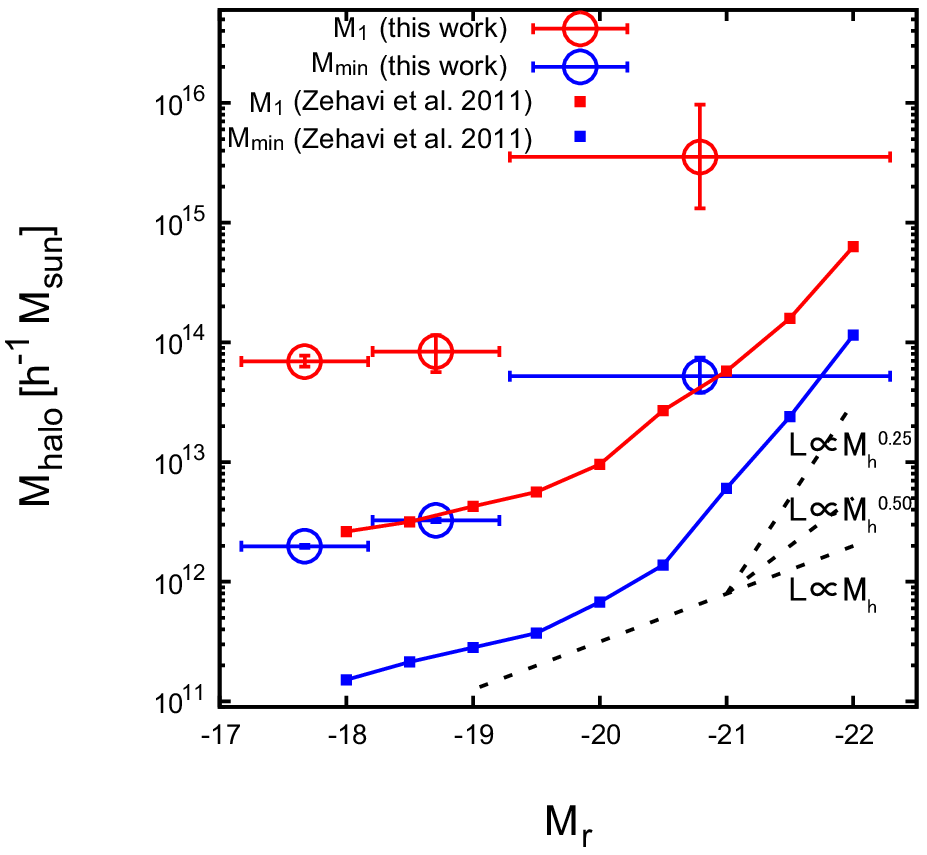}
\caption{The relationship between the threshold absolute magnitudes of sgzK galaxies and the HOD parameters for the characteristic halo mass whereby one dark halo possesses one satellite galaxy, $M_{1}$; and the threshold mass that can possess a galaxy within a dark halo, $\Mmin$. The red points show $M_{1}$, the blue points show $\Mmin$, and the square symbols with lines are the results from \citet{zehavi11}, derived using SDSS galaxy samples. The dotted lines show various relationships between the luminosity of galaxies and the halo mass (i.e., $L \propto M_{{\rm h}}$, $M^{0.5}_{{\rm h}}$, and $M^{0.25}_{{\rm h}}$, from bottom to top). }
\label{fig:Mh-L}
\end{figure}

Figure \ref{fig:Mh-L} shows the relationship between the threshold absolute $K$-band magnitudes of the sgzKs and the HOD parameters $M_{1}$ and $\Mmin$. 
\citet{zehavi05} showed that the halo mass of galaxies with luminosity $L < L_{\ast}$ was weakly dependent on the luminosity of the galaxy, whereas the halo mass of galaxies with a luminosity $L > L_{\ast}$ increased sharply with increasing luminosity. 
This is because a large fraction of baryons within a massive dark halo tend to form satellite galaxies that are below the luminosity threshold, and then accrete to the central galaxy. \citet{zehavi11} also found that the clustering amplitude of galaxies with luminosity $L < L_{\ast}$ increased slowly, whereas the clustering amplitude increased sharply for $L \ga L_{\ast}$, showing that $M_{1}$ and $\Mmin$ had a similar dependence on the luminosity. 
Moreover, \citet{zheng09}, using the same HOD model with varying $\alpha$ and $\sigmaM$, reported a similar relation for luminous red galaxies at $z \sim 0.3$, showing that the dark halo mass of the low-mass halo is proportional to the luminosity of the central galaxy $(L \propto M_{{\rm h}})$, whereas high-mass dark haloes followed the relation $L \propto M_{{\rm h}}^{0.5}$. 

As shown in Figure \ref{fig:Mh-L}, our data show different slopes for the bright sgzKs and faint sgzKs, although the slope of the power law was somewhat uncertain because we have only three data points, which makes it impossible to determine the turning magnitude; this trend is also seen in the local Universe \citep[e.g.,][]{zheng09,zehavi11}. 
Our results for high-mass sgzKs exhibit a similar behaviour to the high-mass low-$z$ galaxies \citep{zehavi05,zehavi11,zheng09}, indicating that baryons are less likely to accrete into the central galaxies at $z \sim 2$, which is also seen for low-$z$ galaxies. 
\citet{zheng09} inferred that baryons are likely to be consumed to form faint satellite galaxies rather than accrete into bright central galaxies. 
Our result of a massive dark halo indicates that the trend may be true, even in the $z \sim 2$ Universe. 

$\Mmin$, representing the threshold dark halo mass of the central galaxy formation, are larger at $z \sim 2$ than $z = 0$. 
This suggests that galaxies at higher-$z$ can be formed only in very massive dark haloes, whereas less-massive dark haloes can form galaxies at lower-$z$, which is known as downsizing \citep[e.g.,][]{fontanot09}. 
On the other hand, $M_{1}$, which represents the threshold dark halo mass of the satellite galaxy formation, are also larger at $z \sim 2$ than $z = 0$. 
Our result implies that satellite galaxies are less likely to be formed in the high-$z$ Universe than in the local Universe; also, the HOD mass parameters show the same luminosity dependence as that for the local Universe. 
It should be noted that the galaxy populations of this work and \citet{zehavi11} are not completely identical; \citet{zehavi11} used volume-limited SDSS galaxy samples ($z \leq 0.2$), whereas our sample is confined to the star-forming galaxy at $z \sim 2$. 

\subsection{The Evolution of sgzKs by Tracing the Halo Mass Evolution}
The mass assembly history can be traced from the mass evolution of dark haloes because the halo mass increases monotonically as a function of cosmic time due to merging. 
The halo mass evolution can be evaluated from the extended Press--Schechter model \citep[EPS model;][]{ps74,bond91,bower91}, which describes the number density of objects in the Universe by modelling the halo merger, which is essential for the $\Lambda$CDM structure formation. 
A more detailed description of the EPS model can be found in \citet[][and the references therein]{hamana06}. 

\begin{figure}
\centering
\includegraphics[width=80mm]{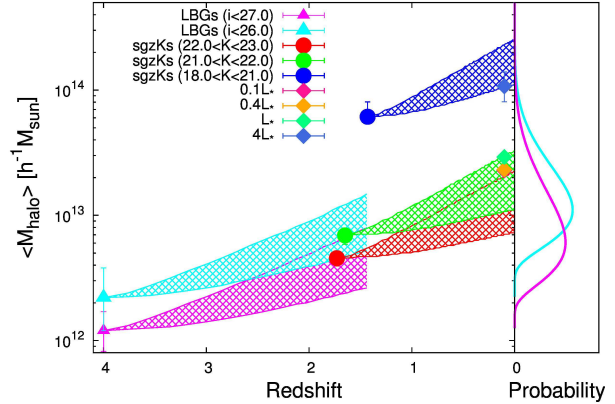}
\caption{Left panel: The mass evolutions of each galaxy population. The masses of the LBGs satisfying $i' < 27.0$ (magenta) and $i' < 26.0$ (cyan) at $z = 4$ are those reported by \citet{hamana06}; the dark halo mass evolutions were calculated using the EPS formalism with a 68$\%$ confidence interval as shown by the hatched regions. Our sgzKs calculated using the HOD formalism are also shown, with the dark halo mass evolutions from $z \sim 2$ to $z \sim 0$. In addition, we show the mean dark halo masses of SDSS galaxies reported by \citet{zehavi11}, which were estimated using their reported HOD parameters. Right panel: The probability distribution functions (PDFs) of mass assembly of the LBGs. The PDF shows the probability of mass evolution of a dark halo that possesses faint LBG (magenta) and bright LBG (cyan) at $z = 4$ to $z \sim 2$. }
\label{fig:EPS}
\end{figure}

First, we trace the dark halo mass evolution of LBGs at $z \sim 4$ derived by \citet{hamana06} to investigate the relationship between sgzKs and the higher-$z$ galaxy population. 
Figure \ref{fig:EPS} shows a comparison of the expected dark halo masses between LBGs and sgzKs. 
It should be noted that the halo occupation models that are adopted by this work and \citet{hamana06} are different. 
The HOD model of \citet{hamana06} is a simple power-law occupation that does not distinguish the central galaxy occupation from the satellite galaxy occupation. 
All assumptions (e.g., cosmological parameters and halo mass function) of \citet{hamana06}, except for $\sigma_{8}$ that was assumed to be $\sigma_{8} = 0.9$ in \citet{hamana06}, are the same as those we adopted. 
LBG samples of \citet{hamana06} were taken from the Subaru/XMM-Newton Deep Survey \citep{ouchi05}. 
The mass evolutions of LBGs from $z = 4$ to $z \sim 2$ are shown in Figure \ref{fig:EPS}, where the hatched regions represent the 68$\%$ confidence intervals of the redshift evolution of the mean dark halo masses of LBGs.
The dark halo mass of our bright sgzKs $(21.0 < K \leq 22.0)$ was equivalent to the evolved dark halo mass of bright LBGs $(i' < 26.0)$ at $3 < z < 4$, whereas the dark halo mass of our faint sgzKs $(22.0 < K \leq 23.0)$ was equivalent to the mass of the evolved dark halo mass of faint LBGs $(i' < 27.0)$, although there was a slight difference between the mass of the bright sgzKs and the faint sgzKs. 
This result may be expected, because galaxies with a higher SFR at larger redshifts produce a considerably greater number of stars and evolve into galaxies with larger stellar mass at smaller redshifts. 
However, we should consider that these two galaxy populations were selected using different selection criteria. 
It should be noted that the dark halo masses of sgzKs with magnitude $21.0 < K \leq 22.0$ and $22.0 < K \leq 23.0$ satisfy the evolved dark halo masses of both bright and faint LBGs within the range of 1$\sigma$ confidence intervals, calculated using the EPS formalism. 
Thus, we cannot clearly determine which LBG is an ancestor of the bright/faint sgzKs; however, the probability distribution functions (PDFs) of mass assembly of LBGs show that the descendants of bright (faint) LBGs at $z \sim 4$ are likely to be bright (faint) sgzKs (see right panel of Figure \ref{fig:EPS}). 
It is necessary to consider the evolutionary link of baryonic properties (e.g., stellar mass and galaxy age), in addition to the evolution of dark halo mass, to validate the evolutionary history of galaxies. 

Figure \ref{fig:EPS} shows that there is a trend whereby the mean halo mass was smaller for LBGs and sgzKs with fainter limiting magnitudes. 
This is caused by different physical mechanisms for each galaxy population. 
For LBGs, the detection band was the $i'$-band, which corresponds to $\sim$ $1500 {\rm \AA}$ in the rest frame. 
LBGs are known to be actively star-forming, and the UV radiation is mainly emitted by young, massive stars; i.e., more-luminous LBGs at UV wavelengths have higher SFRs, which are known to be strongly correlated with $\Mstar$ (``main-sequence'' of the SFR--stellar mass plane) \citep[e.g.,][]{daddi07,oliver10,whitaker12}. 
\citet{hathi13} reported that LBGs up to $z \sim 5$ follow the same correlation between the SFR and the stellar mass, which is consistent with the results of the cosmological hydrodynamical simulation reported by \citet{finlator06}. 
Therefore, the star-formation activity of LBGs was more active with a massive dark halo at $z \sim 4$, and the stellar mass of the galaxies were much higher than UV-faint LBGs that reside in less-massive dark haloes. 

We note that sgzKs were selected according to the $K$-band luminosity, which is contributed to mainly by old, less-massive stars, and corresponds to the stellar mass directly. 
A strong correlation has been reported between the $K$-band luminosity of sBzKs and the stellar mass \citep[e.g.,][]{lin12,bethermin14}. 
However, the correlation between the dark halo mass of sBzKs/sgzKs and the SFRs remains unclear. 
\citet{bethermin14} argued that the dark halo mass of sBzKs increased up to $200 < {\rm SFR}$ $\Msun {\rm yr}^{-1}$ and exhibited a plateau for higher SFRs; however, such flattening of the SFR--halo mass relation is significant only at 1.7$\sigma$; and this should be confirmed via a more precise analysis in the future. 

The expected number of LBGs in a dark halo reported by \citet{hamana04} was $\Ng \sim 0.4$; however, we find the expected number of sgzK to be $\Ng \sim 0.8$. 
This discrepancy may result from differences in the selection criteria. 
The number of satellite galaxies selected by stellar mass at lower-$z$ was more than that selected by SFR at higher-$z$. 
In addition, it was not possible to identify the ancestors of the brightest sgzKs at $z \sim 4$. 
This is because the counterparts of the brightest sgzKs at $z \sim 4$ are thought to be brighter LBGs than those LBGs satisfying $i' < 26.0$; therefore, those bright LBGs are rare objects in the distant Universe. 
Further extensive surveys in the future, i.e., HSC survey, are expected to reveal a sufficient number of such bright LBG candidates to determine the dark halo mass more accurately. 

We traced the expected evolutionary history of our sgzKs to the local Universe. 
Figure \ref{fig:EPS} shows the mass evolutions of sgzKs from $z \sim 2$ to $z \sim 0$ (shown by the hatched regions). 
The mean dark halo masses of local galaxies with various luminosities were calculated by \citet{zehavi11}, and these data were applied to the HOD analysis of local galaxies $(z \leq 0.2)$ from the SDSS DR7 to investigate the dependence of the clustering strength on the physical properties of galaxies (e.g., luminosity and colour). 

From the dark halo mass evolution estimated by the EPS formalism, the faintest sgzKs $(22.0 < K \leq 23.0)$ at $z \sim 2$ $(\Mh \sim 4 \times 10^{12}$ $h^{-1}\Msun)$ evolve to $\Mh \sim (7-20) \times 10^{12}$ $h^{-1} \Msun$, which corresponds to the dark halo mass of the local late-type galaxies, such as the Milky-Way, or typical early-type galaxies. 
Intermediate luminosity sgzKs $(21.0 < K \leq 22.0)$ at $z \sim 2$ with a mass of $\Mh \sim 7 \times 10^{12}$ $h^{-1} \Msun$ evolve to $\Mh \ga 10^{13}$ $h^{-1} \Msun$, which corresponds to the dark halo mass of massive elliptical galaxies or galaxy groups in the local Universe. 
The dark halo mass of the brightest sgzKs $(18.0 \leq K \leq 21.0)$ at $z \sim 2$, with a mass of $\Mh \sim 1 \times 10^{14}$ $h^{-1} \Msun$, evolve to $\Mh \sim (2-4) \times 10^{14}$ $h^{-1} \Msun$, which corresponds to the dark halo mass of the most massive systems in the local Universe, such as rich clusters of galaxies. 

\citet{bethermin14} investigated galaxy evolution by tracing the dark halo mass from $z \sim 2$ to $z \sim 0$ using the halo growth model reported by \citet{fakhouri10}. 
They inferred that sBzKs with a mass of $\Mh \sim 3 \times 10^{11} \Msun$ grew to the most massive field galaxies with a mass of $\Mh \sim 3 \times 10^{12} \Msun$, a halo with a mass of $\Mh \sim 3 \times 10^{12} \Msun$ grew to form galaxy groups, and sBzKs/pBzKs with a mass of $\Mh \sim 3 \times 10^{13} \Msun$ evolved into galaxy clusters in the local Universe. 
This differs slightly from our analysis; we find that the halo mass evolution was approximately $\sim 0.5$ dex from $z \sim 2$ to $z \sim 0$, whereas the halo mass evolution reported by \citet{bethermin14} was approximately one order of magnitude higher. 
This difference in the halo growth rate led to differences in the corresponding progenitors of the $z \sim 2$ galaxies. 

In summary, our results suggest that faint LBGs at $z \sim 4$ could evolve into faint sgzKs $(22.0 < K \leq 23.0)$ at $z \sim 2$ and to the Milky-Way-like galaxies or elliptical galaxies in the local Universe, whereas bright LBGs at $z \sim 4$ could evolve into intermediate luminosity sgzKs $(21.0 < K \leq 22.0)$ and into the most-massive elliptical galaxies or central galaxies of the galaxy groups in the local Universe, and the most-luminous sgzKs $(18.0 \leq K \leq 21.0)$ could evolve into the most massive systems in the local Universe; i.e., the central galaxies of the galaxy clusters. 
However, these evolutionary scenarios were determined by only considering the evolution of the dark halo mass. 
It may not be straightforward to relate different galaxy populations at different redshifts, as they may follow different star-formation histories, which are traced by other physical properties such as SFR, stellar mass, and age. 
In the following section, we discuss the evolution of galaxy populations, which provides another point of view from which to analyse our galaxy evolution model. 

\subsection{Galaxy Evolution by Tracing the Number of Satellite Galaxies}
One advantage of the HOD formalism is that it enables us to describe the number of galaxies in a dark halo for a given dark halo mass. 
Because of this characteristic of the HOD formalism, we were able to trace the number evolution of satellite galaxies within a dark halo, which is not possible if we assume one-to-one galaxy--halo correspondence. 
In this section, we discuss the number evolution of satellite galaxies in a dark halo using our sgzKs, assuming that galaxies evolve so as to follow the dark matter halo evolution predicted using the EPS (see Sec.~5.1). 

First, we consider the evolution of dark halo with $M'_{1}$; this defines the mass of the dark halo occupied by only one galaxy at $z \sim 2$, satisfying $N(M'_{1})=1$ for each luminosity subsample, to investigate the number evolution of the satellite galaxy within the dark halo. 
We follow the evolution of the dark halo with mass of $M'_{1}$ based on the EPS formalism from $z \sim 2$ to $z = 0$; the dark halo mass at $z =0$ was determined as the mass in which the PDF of the dark halo mass evolution reaches a peak; its error is in the range of 1$\sigma$ confidence interval of the PDF. 
The number of the satellite galaxies in the dark halo at $z = 0$ was calculated by the halo occupation function (equation \ref{eq:N_gal}) at an evolved dark halo mass, with the best-fit HOD parameters of \citet{zehavi11}. 
We note that we took into account both errors of the evolved dark halo mass and HOD parameters of \citet{zehavi11} to evaluate the number of satellite galaxies. 

\begin{figure}
\centering
\includegraphics[width=80mm]{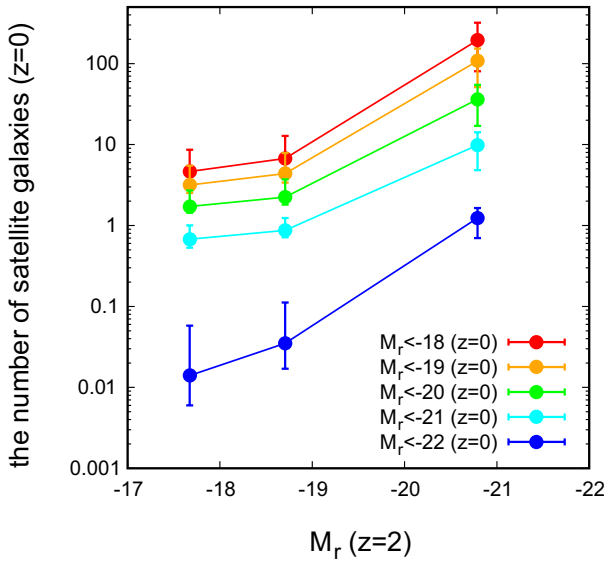}
\caption{The evolution of the number of satellite galaxies in the dark halo. The horizontal axis is the magnitude threshold of the absolute magnitude of sgzKs at $z \sim 2$, and the vertical axis is the number of satellite galaxies in the dark halo at $z \sim 0$, which contains only one sgzK satisfying each limiting magnitude. The different colours of the data points correspond to differences in the magnitude threshold of the satellite galaxies.}
\label{fig:sate}
\end{figure}

Figure \ref{fig:sate} shows the expected number of satellite galaxies at $z = 0$ for a dark halo, which corresponds to a central galaxy with no satellite galaxies at $z \sim 2$. 
In this figure, the horizontal axis represents the absolute rest $r$-band magnitudes, which corresponds to the $K$-band limiting magnitude of the sgzKs at $z \sim 2$, and the vertical axis is the number of satellite galaxies that are expected to be formed in the dark haloes at $z = 0$. 
The colour difference indicates a different absolute magnitude of satellite galaxies at $z = 0$. 
More-luminous galaxies tend to have more satellite galaxies at $z = 0$, and a large fraction of these satellite galaxies are faint. 
In dark haloes that contain the faintest central sgzKs with $M_{\rm r} \ga -19$ at $z \sim 2$, approximately three satellite galaxies with luminosity equal to that of central sgzKs at $z \sim 2$ are expected to form.  
However, brighter satellite galaxies are unlikely to form in the dark haloes of these faint sgzKs. 
In other words, dark haloes with the faintest sgzK evolve into dark haloes that contain $ \la 10$ faint galaxies and no bright galaxies at $z = 0$. 
In the dark haloes that contained the brightest central sgzKs $M_{\rm r} \ga -22$ at $z \sim 2$, approximately $100$ satellite galaxies are expected to be formed by $z = 0$. 
This is consistent with the results described in Section~5.2; i.e., local galaxy groups/clusters that contain 100 or more member galaxies. 
Few of the brightest satellite galaxies had similar luminosity to those of sgzKs; i.e., sgzKs that reside in haloes at $z \sim 2$ would become a central galaxy of a galaxy cluster in the local Universe, especially the brightest cluster galaxies (BCGs). 

Based on the discussion of the number evolution of satellite galaxies, the faintest sgzKs appear to evolve into the Milky-Way-like galaxies whereas the brightest sgzKs evolve into the central galaxies of galaxy clusters in the local Universe, especially BCGs. 
These results are consistent with the implications of the discussion of galaxy evolution from considerations of the evolution of the dark halo mass. 

\subsection{Stellar-to-Halo Mass Ratio}
\begin{figure}
\centering
\includegraphics[width=80mm]{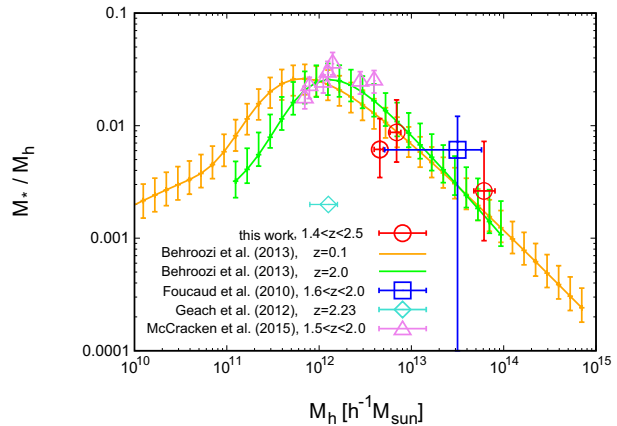}
\caption{A comparison of SHMRs between our results for sgzKs (red) and the theoretical predictions reported by \citet{behroozi13}. The SHMR reported by \citet{foucaud10} (blue), \citet{geach12} (cyan), and \citet{mccracken15} (purple) are also shown. The calculation method used to obtain our dark halo masses was HOD analysis, and the stellar masses were found using the $(z-K)$ colour and $K$-band luminosities of the sgzKs. \citet{geach12} and \citet{mccracken15} also calculated dark halo masses using HOD analysis, whereas the method used by \citet{foucaud10} was large-scale galaxy clustering. The stellar mass of all of those studies was calculated by SED fitting. The solid curves with error bars show results of the theoretical model reported by \citet{behroozi13}. }
\label{fig:SHMR}
\end{figure}

Figure \ref{fig:SHMR} shows a comparison between the results of our SHMRs, the theoretical prediction of \citet{behroozi13}, and the observed SHMRs reported by \citet{foucaud10}, \citet{geach12} and \citet{mccracken15}. 
The stellar mass was determined from the $K$-band magnitudes and $(z - K)$ colours of each of the sgzK galaxies \citep[][see paper~I for more details]{kodama04,koyama13}, and $\Mh$ was used as the dark halo masses. 
It should be noted that the SHMR, in principle, should be calculated by the ratio between the dark halo mass and all stellar components within the dark halo \citep[e.g.,][]{kravtsov14}; however, \citet{behroozi13} calculated the SHMRs via an abundance matching method. 
Our data, as well as those of \citet{geach12} and \citet{mccracken15}, were calculated using an HOD analysis. 
Our data points are almost in good agreement with the theoretical model, as well as the other observations. 
The SHMRs of our sgzKs exhibit almost identical values at $z \sim 0$ within the $1 \sigma$ confidence level, consistent with the theoretical predictions for SHMRs with $M_{{\rm h}} \ga 10^{12}$ $h^{-1} \Msun$, and do not significantly change between $z \sim 0$ and $z \sim 2$. 

Our results are almost consistent with the prediction of the model reported by \citet{behroozi13}; however, there was a slight difference in the trend, especially for the SHMR of the least massive haloes; the SHMR calculated from our data was smaller than the prediction of the model at $z \sim 0$ as well as at $z \sim 2$. 
This discrepancy may be due to differences in the galaxy populations. 
Our data were based on sgzKs, whereas \citet{behroozi13} did not take into account galaxy types when assigning galaxies to dark haloes via abundance matching. 
\citet{tinker13} also investigated this relationship over the redshift range of $0.22 < z < 1.00$, distinguishing the differences in galaxy populations as star-forming galaxy samples, passive galaxy samples, and all galaxy samples. 
They inferred that, especially for galaxies with a halo mass of $M_{{\rm h}} \ga 10^{12}$ $\Msun$, the stellar mass of a central galaxy depends on the population of the galaxy samples with a fixed dark halo mass due to differences in the growth rates of galaxies. 
With our samples, however, we consider only star-forming galaxies, whereby the central galaxy has a smaller stellar mass than the other two galaxy populations \citep{tinker13}. 
This may lead to a smaller SHMR at $M_{{\rm h}} \sim 3 \times 10^{12}$ $h^{-1} \Msun$ than that reported by \citet{behroozi13}. 
The  smaller SHMR of our data compared with that of \citet{mccracken15} that sampled all galaxies, including passive galaxies, may also be attributed to differences in the galaxy populations. 
The HAEs \citep{geach12}, which are in the early phase of star-forming activity, also show significantly lower SHMR than other galaxies. 

\section{CONCLUSIONS}
We have described the results of HOD analysis of star-forming galaxies at $z \sim 2$. 
Our high-quality ACFs were based on a large sgzK galaxy sample and are sufficiently accurate to allow the use of HOD analysis to determine accurate dark halo masses. 

The major conclusions of this work may be summarized as follows. 

\begin{enumerate}
\item  The ACFs of sgzKs were well described by the halo model based on the HOD formalism, and we were able to estimate accurate dark halo masses and HOD parameters. 
The resulting HOD parameters enabled us to determine the dark halo mass, $\Mh$, and the expectation number for dark haloes. 
The mean dark halo masses of our sgzK galaxies were $(4.55^{+0.42}_{-0.60}) \times 10^{12}$ $h^{-1} \Msun$, $(6.91^{+0.77}_{-1.05}) \times 10^{12}$ $h^{-1} \Msun$, and $(6.12^{+1.92}_{-1.37}) \times 10^{13}$ $h^{-1} \Msun$ for $22.0 < K \leq 23.0$, $21.0 < K \leq 22.0$ and $18.0 \leq K \leq 21.0$, respectively. 
The HOD mass parameters, $M_{1}$, $\Mmin$, $M_{0}$, and $\Mh$, almost all increased monotonically as a function of the magnitude threshold, which suggests that more-luminous galaxies reside in more-massive dark haloes. 
The expectation numbers for sgzKs in a dark halo were nearly consistent within the different luminosity bins. 
This implies that the expectation number of faint sgzKs residing in less-massive dark haloes and bright sgzKs residing in more-massive haloes was almost identical. 
The HOD-based mean dark halo masses were several times larger than the mean halo masses determined by large-scale galaxy clustering. 
This discrepancy may be due to differences in the definition of the mean halo mass, which was also suggested by \citet{geach12}. 

\item We find that $M_{1}$ and $\Mmin$ for faint sgzKs are approximately proportional to the luminosity, whereas bright sgzKs follow $L \propto M_{{\rm h}}^{0.5}$, as is the case for galaxies in the local Universe. 
This is because baryons in massive dark haloes tend to form faint satellite galaxies, which are below the luminosity threshold, rather than accrete central galaxies; this confirms the same trend at $z \sim 2$. 
Our $\Mmin$ was larger than in the local Universe \citep{zehavi11} over all magnitudes, implying that galaxies are formed only in massive dark haloes at $z \sim 2$, and that less-massive dark haloes form galaxies more slowly, known as downsizing. 
Additionally, $M_{1}$ was also larger than that of \citet{zehavi11}, suggesting that satellite galaxies are less likely to be formed at $z \sim 2$ than in the local Universe. 

\item We investigated the relationship between sgzKs and galaxy populations at low- and high-$z$ by tracing the dark halo mass using the EPS formalism. 
Faint LBGs $(i' < 27.0)$ at $z \sim 4$ are expected to evolve into faint sgzKs $(22.0 <K \leq 23.0)$, whereas bright LBGs $(i' < 26.0)$ are expected to evolve into intermediate luminosity sgzKs $(21.0 < K \leq 22.0)$. 
The brightest sgzKs at $z \sim 2$ appear to have evolved from more luminous LBGs, the dark halo mass of which is difficult to calculate due to the small number density. 
We expect to be able to obtain further luminous LBG candidates for the ancestors of our brightest sgzKs in future wide-field observation. 
We also considered the evolution of the halo mass of our sgzKs to the local Universe and compared this with the results of \citet{zehavi11}.
Our analysis suggested that faint sgzKs could evolve into Milky-Way-like galaxies or elliptical galaxies, intermediate luminosity sgzKs could evolve into massive elliptical galaxies or the central galaxies of galaxy groups, and that the brightest sgzKs could evolve into the central galaxies of galaxy clusters; i.e., BCGs in the local Universe. 

\item We also investigated the number evolution of satellite galaxies in a dark halo using HOD analysis. 
In a dark halo that contains one faint sgzK at $z \sim 2$, $\la 10$ faint satellite galaxies with similar luminosity to the central sgzK were formed at $z = 0$. 
However, a dark halo that contained only one of the brightest sgzKs evolves into a massive system that contains more than 100 galaxies in the local Universe, which corresponds to galaxy clusters. 
This is consistent with the results of galaxy evolution based on the evolution of the dark halo mass. 

\item We calculated the SHMRs from our sgzKs, and compared these with predictions using the model reported by \citet{behroozi13}. 
Our results were in good agreement with the results of the model; i.e., the SHMR at $z \sim 2$ did not change significantly compared with the equivalent relation at $z \sim 0$; however, our SHMR for the faintest bin was slightly smaller than that reported by \citet{behroozi13} and \citet{mccracken15}. 
One possible explanation for this discrepancy is the difference in galaxy populations. 
\end{enumerate}

\section*{acknowledgments}
We thank the referee for his/her useful comments to improve this paper. 
This work is based on data collected at the Subaru Telescope, which is operated by the National Astronomical Observatory of Japan (NAOJ). 
This research was supported by the Japan Society for the Promotion of Science through Grant-in-Aid for Scientific Research 23340050 and 15H03645. 
This work was supported in part by the Center for the Promotion of Integrated Sciences (CPIS) of SOKENDAI. 

This research is based in part on data obtained as part of the United Kingdom Infra-Red Telescope (UKIRT) Deep Sky Survey. 
This study is based on observations obtained with MegaPrime/MegaCam, a joint project of CFHT and CEA/DAPNIA, at the Canada--France--Hawaii Telescope (CFHT), which is operated by the National Research Council (NRC) of Canada, the Institut National des Sciences de l'Univers of the Centre National de la Recherche Scientifique (CNRS) of France, and the University of Hawaii. 
This work is based in part on data products produced at TERAPIX and the Canadian Astronomy Data Center as part of the Canada--France--Hawaii Telescope Legacy Survey, a collaborative project of NRC and CNRS.

\end{document}